# High-$T_c$ Superconductivity: Strong Indication of Filamentary-Chaotic Conductance and Possible Routes to Superconductivity Above Room Temperature.

**Hans Hermann Otto**

Materialwissenschaftliche Kristallographie, TU Clausthal, Adolph-Roemer-Str. 2A,
D-38678 Clausthal-Zellerfeld, Germany
Email: hhermann.otto@web.de

**Abstract**

The empirical relation of $T_{co}(K) = 2740/<q_c>^4$ between the transition temperature of optimum doped superconductors $T_{co}$ and the mean cationic charge $<q>_c$, a physical paradox, can be recast to strongly support fractal theories of high-$T_c$ superconductors, thereby applying the finding that the optimum hole concentration of $h^+ = 0.229$ can be linked with the universal fractal constant $\delta_1 = 8.72109…$ of the renormalized *Hénon* map. The transition temperature obviously increases steeply with a domain structure of ever narrower size, characterized by *Fibonacci* numbers. With this backing superconductivity above room temperature can be conceived for synthetic sandwich structures of $<q>_c$ less than 2+. For instance, composites of tenorite and cuprite respectively tenorite and CuI (CuBr, CuCl) onto AuCu alloys are proposed. This specification is suggested by previously described filamentary superconductivity of 'bulk' $CuO_{1-x}$ samples. In addition, cesium substitution in the Tl-1223 compound is an option. A low mean cationic charge allows the development of a frustrated nano-sized fractal structure of possibly ferroelastic nature delivering nano-channels for very fast charge transport, in common for both high-$T_c$ superconductor and organic inorganic halide perovskite solar materials.

## 1. Introduction

The recent discovery of conventional superconductivity at the highest until today known transition temperature of 190 K on hydrogen sulfide at a high pressure >150 GPa by *Drozdov, Eremets and Troyan* (2014) [1] gives rise to discuss again a possible route to superconductivity above room temperature. However, high pressure is not recommended for everyday use. Very recently, the crystal structure of this high pressure modification of $H_2S$ was solved, indicating an anti-perovskite structure in the sense that $SH^-$ represents the A site and $SH_3^+$ the B site of this structure type [2]. Below I will suggest a hypothetical anti-perovskite compound that would, if chemically accessible, a highly interesting option for a room temperature superconductor. Some years ago we published an experiment, which indicated filamentary, but not stable superconductivity at 220 K of an oriented multi-phase sample of the Y-Ba-Cu-O system, deposited on a (110)-$SrTiO_3$ substrate (*Schönberger et al.*, 1991) [3]. The uncommon (110) orientation of the substrate surface was chosen to provoke symmetry reduction and ferroelastic domain formation of the deposited thin film phases by strain. We identified the minor compounds $BaCuO_2$ and CuO, respectively, besides the superconducting main phase $YBa_2Cu_3O_{7-\delta}$. A comparable result with a large resistivity drop at 220 K has been published little earlier by *Azzoni et al.* (1990) [4] on reduced cupric oxide samples, which showed some red $Cu_2O$ besides CuO. Later *Osipov et al.* (2001) [5] deposited Cu films onto cleaned CuO single crystal surfaces and observed at the CuO-Cu interface a giant electric conductivity increase by a factor up to $1.5 \cdot 10^5$ even at 300 K. Today it is assumed that the observed rapid drop of the electric conductivity at 220 K and 300 K is caused by superconductivity of oxygen deficient $CuO_{1-\delta}$ filaments [3][4][5].

Consequently, this finding has been patented, by others, as ultra-conductor of bundled filaments of copper oxide coated copper wires and foils, respectively [6].

Interestingly, artificial interfaces between insolating perovskites that indicate superconducting response are described in 2007 by Reyen et al. [7]. Further, I quote an investigation published recently by Rhim et al. (2015) [8] regarding possible superconductivity via excitonic pairing onto an interfacial structure between CuCl and Si(111), remembering that electronic anomalies in cuprous chloride have been described by Chu et al. [9] long time ago. Both publications encouraged me even more to put earlier ideas about copper oxide composites on record.

A hypothetical $BaCuO_2$ phase with puckered T-CuO nets, in contrast to planar $CuO_2$ nets of the $BaCuO_2$ infinite layer phase, was recently proposed by the present author as an alternative high-$T_c$ candidate together with a large cupric oxide cluster compound [10]. However, sandwich structures composed of tenorite and cuprite as well as tenorite and CuI respectively CuBr or CuCl are more interesting. Also cesium substitution in the Tl-1223 compound is a route to reach a high transition temperature. The composition for such composites will be passed by the empirical relation of $T_{co} = 2740/<q>_c^4$ that connects the transition temperature $T_{co}$ of optimum doped superconductors with the mean cationic charge $<q>_c$ [11]. Moreover, this paradox relation will be traced back to a physically more accepted basis. Both, the theoretical results and the practical suggestions regarding T-CuO nets with twice as many copper atoms as in the $CuO_2$ nets should envisage physicists to realize a dream of mankind, superconductivity above room temperature.

This publication was almost completed just as the message about an apparently successful synthesis of a room temperature superconductor (with far withheld information about chemical composition!) astonished the scientific community [12]. Therefore, some of my results may already be outdated, what I don't hope yet, because nature never puts all eggs in one basket.

## 2. Results and Discussion

The most abounded mineral in Earth is $MgSiO_3$, a perovskite phase confined to the high pressure region of the Earth mantle, but also discovered in a shocked meteorite and now named bridgmanite [13]. The dominance of perovskites and derived phases continues in our technology determined life due to their wide variety of adapted symmetries and the surprising diversity of physical properties such as ferrolectricity, ferroelasticity, superconductivity, ionic conductivity, photocatalytic property and efficient photovoltaic respond, respectively. Some of these properties are mutually dependent.

Beginning with the last mentioned challenge, the development of efficient solar cells, you just witnessed the great progress in organic inorganic halide perovskite solar cells. Their very high solar efficiency is caused by the low cationic charge of $<q>_c = 1.5^+$, resulting in a low and less confining *Madelung* potential that favors a small-sized ferroelastic and conductive domain structure with enhanced charge separation and improved carrier lifetime [14].

## 2.1 Resolving a Paradox: $T_{co}$ (K) = 2740/$<q_c>^4$.

A less restricted freedom of electronic movement may also be suggested for the family of superconductors related to oxide perovskites. The rise of the transition temperature $T_{co}$ goes inversely proportional with the fourth power of the mean cationic charge $<q>_c$. Figure 1 depicts $<q>_c$ versus $T_{co}$, somewhat simplified compared to previous results [11] and now branched into electron (conventional) conductors and hole (unconventional) conductors. The more complicated relation given in the earlier paper was induced by the idea to cover both conventional and unconventional superconductors and by the assumption that a mean cationic charge of 2+ would be a natural crystal-chemical limit. Remarkably, the data summarized in that reference show also exemplarily pnictide-based superconductors. The data given in reference [11] were now supplemented with the red-marked samples, especially including $CuO_{1-x}$ as hole conductor, and with the result for $H_2S$ under high pressure as electron conductor. The curve for electron transport will be discussed in a forthcoming contribution. The magenta colored curve for hole superconductors (Figure 1) is represented by the relation

$$T_{co} (K) = 2740/<q_c>^4. \tag{1}$$

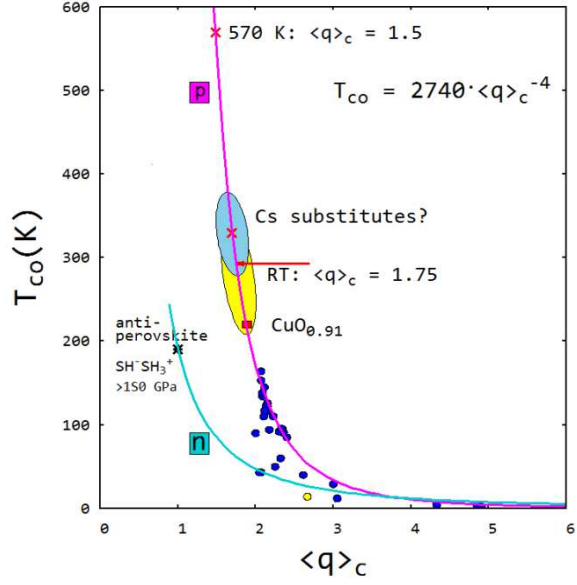

**Figure 1**. Mean cationic charge $<q>_c$ versus transition temperature $T_{co}$ (K) of optimal doped superconductors (see also [11]). Branch of n-type superconductors turquoise, p-type ones magenta. Yellow circle spinel phase superconductor, colored ellipses: composites explained in the text.

Surprisingly, $<q>_c \approx 1.5^+$ emerges as high-$T_{co}$ asymptotic limit. With a successive reduction of the mean cationic charge in direction of $<q>_c = 1.75$ one would reaches the field of possible room temperature superconductivity. Recently, the author proposed hypothetical compound, a cagy cuprate cluster compound as well as tetragonal BaCuO$_2$ with puckered T-CuO nets [10]. Both compounds could reach a transition temperature around 180 K, if their synthesis with sufficient doping would be successful.

The scale factor of Eq. 1 of $s_c = 2740$ K can intuitively traced back to give

$$s_c = 8\varepsilon^2 m_e \frac{v_F^4}{v_K^2} \frac{1}{k_B} = 2736 \text{ K} \tag{2}$$

where $v_K = \frac{e^2}{\varepsilon_0 h} = 4.37538 \cdot 10^6$ m/s is a universal constant with the dimension of a speed, $e$ the elementary charge, $h$ the Planck constant, $v_F$ the *Fermi* speed, $\varepsilon$ the permittivity of the medium, $\varepsilon_0$ the permittivity of free space, and $m_e$ the electron mass, respectively. With $\varepsilon = 5.04$ and an assumed *Fermi* speed of $v_F = 2.5 \cdot 10^5$ m/s, respectively, the scaling factor agrees well with the fitted parameter of 2740 K. If a higher effective mass of the quasiparticles is chosen, for instance $m_h \approx 1.5 m_e$, then the *Fermi* velocity has to be reduced correspondingly, as was observed experimentally in the region of optimal doping. With $\tilde{v}^2 = v_F^4 / v_K^2$ the result yields

$$k_B T_{co} = \frac{16\varepsilon^2}{<q>^4} \frac{1}{2} m \tilde{v}^2 \ . \tag{3}$$

This represents rather the total energy of a system of oscillating charged quasiparticles, dispersed in a medium of permittivity $\varepsilon$, than solely coulomb energy. The dimensionless variable $<q>_c$, with the fourth power instead of the expected second one (in three dimensions of space), obviously subsumes, beside the cationic charges and the hole concentration, also camouflaged anionic contributions and the effect of steadily increasing permittivity, because the compounds of highest $T_{co}$ are composed of highly polarizable heavy atom ions. How the matter might be, the relation between $T_{co}$ and $<q>_c$ can serve to predict possible critical temperatures solely from charges of a compound. Besides a low cationic charge a high permittivity (indicating certain proximity to phase transitions) may favorable to support high $T_{co}$.

Some ideas should be passed on that are related to optimum values of the hole concentration found, and to obvious accumulations in certain $T_{co}$ values. Beginning with the value of optimum $h^+$ of $\sigma_o = 0.229$ (or multiples) [15] that is near 3/13 and is attributed to a large group of high-$T_c$ superconductors based on Tl or Hg, respectively. Surprisingly, the multiplier, which would give two holes needed to create a pair, emerges as the number

$$\boxed{\delta_1 = 8.721}$$

known as a universal scaling constant for two-dimensional maps in the theory of fractal systems or chaotic ones, with the precise value of $\delta_1 = 8.7210972…$[16] [17]. Recently, *Savin et al.* [18] studied the self-oscillating system of the *Van der Pol* oscillator [19] subjected to an external force to compensate dissipation. Scaling constants $\delta_1$ (and $\delta_2 = 2$) have been determined as eigenvalues of the matrix containing the existence intervals of two subsequent cycles of the periodic-doubling cascade in the parameterized version of a Hénon map with renormalized (x,y)-parameters.

The optimum concentration of holes is given per cuprate layer per unit-cell. Maintaining two dimensions along cuprate layers, then two electrons reside in a slab of the extension $\delta_1 a^2 = \pi l^2 a^2$. The 'domain' extension multiplier is then given by the quotient of *Fibonacci* numbers

$$l = \sqrt{(\delta_1/\pi)} = 5/3.$$

Pnictide superconductors yield an optimum hole concentration of about $\sigma_o/4$ [15] and can be traced back to $\delta_1$ too. Further, although it is already difficult to identify optimum doping, picking out accumulations in the $T_{co}$ ranking is the more difficult. Nevertheless, one can try it as *Mitin* [20] has done before, identifying such $T_{co}$ groups with domain widths. Following the notation of that author, he obtained for the transition temperature $T_{co}$ related to assumed zig-zag bosonic stripe domains of width $w = \eta \cdot a$, connected with oxygen interstitials,

$$T_{co} = \frac{h^2}{32\pi^2 m_e r_o^2 k_B (2\eta^2+\eta)} \tag{5}$$

where $h$ represents the *Planck* constant, $k_B$ the *Boltzmann* constant, $m_e$ the rest mass of the electron, $a$ an elementary length related to neighboring cations, $r_o = 2.72$ Å a string distance along O–O bonds, and $f(\eta) = \sqrt{(2\eta^2+\eta)}$ a function of the domain extension (rank) $\eta$ with respect to $a$, respectively.

Going a step further, one can associate such $T_{co}$ clumps with *Fibonacci* numbers $f_i$ determining the extension of domains. Possible mixed domain states can be described by the mean of consecutive $f_i$'s. Table 3 compares *Mitin*'s results with my *Fibonacci* number explanation. The results may give a hint to a possible filamentary-chaotic origin of high-$T_c$ superconductivity. Using *Fibonacci* numbers $f_i$ one can roughly formulate (see Table 4):

$$T_{co} \approx 12000/f_i \text{ (K)}. \tag{6}$$

Upon substitution of Eq. 1 into Eq. 6 one gets

$$<q>_c^4 = 0.228 \cdot f_i \approx 2f_i/\delta_1. \tag{7}$$

Surprisingly, the emerging factor corresponds numerically to the charge of $\sigma_o = 0.228$ [15]. It is remarkable that also the quotient of the *Fermi* speed chosen (which is near to reality) and the *Klitzing* speed yields

$$v_F/v_K \approx 0.0571 \approx 1/(4\ \sigma_o) \approx 1/(2\ \delta_1). \tag{8}$$

With this result one can finally express the energy in different and scale-free forms:

$$k_B T_{co} \approx \frac{8\ \varepsilon^2}{f_i \sigma_o} m_e \tilde{v}^2. \qquad k_B T_{co} \approx \frac{\varepsilon^2}{f_i \delta_1} m_e v_F^2. \qquad k_B T_{co} \approx \frac{\varepsilon^2}{4 f_i \delta_1^3} m_e v_K^2. \tag{9}$$

When using the conductance quantum $G_o = 2e^2/h = 7.748091 \cdot 10^{-5}$ (S) one can write down the energy as

$$k_B T_{co} \approx \frac{\varepsilon^2}{16\varepsilon_o^2 f_i \delta_1^3} m_e G_o^2 \tag{10}$$

The multiple energy representation is given by the special nesting property of fractals. Resolving the paradox finally leads to the result that high-$T_c$ superconductors obey fractal conductance behavior, which is intrinsically accompanied with self-similarity and scale-free characteristics. This finding of possible fractality and self-similarity of high-$T_c$ superconductors supports the seminal investigations of *Fratini et al.* [21] on the fractal organization of oxygen interstitials and the recent contribution of *Poccia et al.* [22]. Later *Phillabaum et al.* [23] reports on the fractal structure of nanoscale electron lines on the surface of superconductors that spread out in the bulk. In this context I quote also the zig-zag filamentary theory of high-$T_c$ superconductors of *Phillips* [24] [25].

**Table 5**. Dependence of the superconducting transition temperatures $T_{co}$ on the domain width $w = \eta \cdot a$, $f(\eta)^2 = 2\eta^2+\eta$, where $\eta$ represents the rank introduced by Mitin [20], $f_i$ = Fibonacci numbers.

| $\eta$ | $d_i$ | | Mean of consecutive $f_i$'s | $T_{co}$ | $T_{co} \cdot d_i$ | $<q>_c$ |
|---|---|---|---|---|---|---|
| | $2\eta^2+\eta$ | $f_i$ | | | | |
| | | 8 | | (1500) | 12000 | - |
| 2 | 10 | | 10.5 | (1200) | 12000 | - |
| | | 13 | | (923) | 12000 | - |
| 3 | 21 | 21 | | **570** | 11970 | 1.50 |
| 4 | | 34 | | **353** | 12000 | 1.67 |
| | 36 | | | **334** | 12024 | 1.70 |
| 5 | 55 | 55 | | **219** | 12045 | 1.87 |
| 6 | | | 72 | **167** | 12024 | 2.01 |
| | 78 | | | 155*) | 12090 | 2.05 |
| | | 89 | | **135** | 12015 | 2.12 |
| 7 | 105 | | | **115** | 12075 | 2.21 |
| | | | 116.5 | 103 | 12000 | 2.26 |
| 8 | 136 | | | 89 | 12104 | 2.35 |
| | | 144 | | 84 | 12096 | 2.39 |
| 9 | 171 | | | 70 | 11970 | 2.50 |
| | | | 188.5 | 64 | 12064 | 2.55 |
| 10 | 210 | | | 57 | 12000 | 2.63 |
| | | 233 | | 52 | 12116 | 2.70 |
| 11 | 253 | | | 47 | 12000 | 2.76 |
| 12 | **300** | | 305 | **40** | 12000 | 2.88 |

*) high pressure applied

Recently, *Mushkolaj* [26] compared two mutually inverse $T_c$ functions related to an elastic atom collision model in contrast to an elastic spring model, respectively, given by

$$T_c \text{(collision)} \propto z^{-1} \quad \text{versus} \quad T_c \text{(spring)} \propto z, \tag{9}$$

where $z = (\sqrt{M_1 M_2})\Delta x^2$, and $M_i$ = atom or electron masses, $\Delta x$ = distances.

What would happen, if both interactions compete? Does one have the case of a double pendulum running to chaotic motion states? Solutions of the rational function $f(z) = z - z^{-1}$, which may describe such competition between both underlying physical processes, would point indeed to chaotic behavior of the excited carriers.

A majority of high-$T_c$ superconducting compounds shows a pronounced deviation from the tetragonal symmetry. This supports the formation of ferroelastic domains with walls as charge carrier sink. If the itinerant holes formed by doping travel to the walls, it results at the first moment a stack of positively charged walls separating insulating regions. The repulsive forces have to keep in balance and may strengthen orthorhombicity. However, when the concentration of holes exceeds a certain limit, then repulsive forces will be reduces by the formation of bosons, confining preformed pairs of holes in one-dimensional bosonic stripes [27]. One might expect that lowering of the effective cationic charge in the region between the stripes would favor undulation of the stripes, finally leading to the superconducting state due to their interaction.

However, in which way interstitial oxygen ions could be involved in the pairing scenario? A possible modified exciton pairing mechanism may be as follows. In the vicinity of the large and highly polarizable $Ba^{2+}$ ions the interstitials may show a 'chameleon' feature, being once -1 charged and then again -2, respectively. It means that this oxygen atom is able to easily expel or souk a charge carrier. Apical oxygen atoms on the interlayer between cuprate layers and spacer layer can be involved in this process. Just when the apical oxygen atom is displaced towards the Cu center, enough space is provided to create the larger $O^{1-}$ ion. An exciton may bind a just expelled (hot) hole to form a metastable three-particle-entity that would rapidly decay, leaving a boson and the mediating electron. It needs only few electrons to form a cascade of bosons, and the probability it happens is higher than for an exciton-exciton process. The pairing process can take place locally near interstitial oxygen atoms or within the strain field of ferroelastic domain wall sinks, involving strain effect as important too.

The given ideas about fractality of high-$T_c$ superconductors and a modified pairing mechanics, respectively, should be considered in a modified and more founded (string?) theory. Until then we can work with the auxiliary variable $<q>_c$ as a tool to find candidate systems for superconductivity above room temperature, besides taking care of a high permittivity of the compounds. This is accomplished in the next chapter.

## 2.2 Routes to Superconductivity Above Room Temperature

First, the $T_{co}$ versus mean cationic relation will be applied to make progress with known Tl-based superconductors with $CuO_2$-based nets. Following this, tenorite-based composites are prospected in detail with twice as many copper atoms per layer.

### 2.2.1 A Cs-Substituted Tl-1223 Compound

In short, a resume of the crystal-chemistry of Tl-based superconductors will be given. Crystal-chemical data from own structure determinations on Tl-based cuprates were summarized in Table 1 [28][29] [30]. The reduced electron density found at the Tl position of TlO-double layer compounds is recast in partial occupation by $Cu^{1+}$, and the enhanced value at the Ca position was attributed to a partial $Tl^{3+}$ substitution. The $Cu^{1+}$ content in the calculated amount was confirmed by electron energy loss spectroscopy (*EELS*) on single crystals [31]. What happens, if cations such as $Hg^{2+}$ or $Tl^{3+}$ attain the lower oxidation state with lone electron pairs and associated dipole momentum? The properties of $Bi^{3+}$ are dominated by both the space that the lone electron pair needed and the resulting dipole momentum. Therefore, $Bi^{3+}$ will be shared in *Aurivillius* double-layers, triple domains or channels to minimize the dipole momentum. Indeed, only double layer Bi-based superconductors are found. In the case of $Hg^{1+}$ two ions join to form $Hg_2^{2+}$ with zero resulting dipole momentum, but this was not observed in Hg-based superconductors. If one considers partial replacement of $Tl^{3+}$ by $Tl^{1+}$ in high-$T_c$ superconductors, again one is faced with the effect of the dipole momentum of $Tl^{1+}$, especially if mono-layer compounds such as Tl-1223 are considered. The replacement should be significantly less than halve the Tl content as observed, but may be restricted to the Tl site and not to the Ca site substitution. Because Tl-1223, given as example in Table 1, seemed to be over-doped according to our results, a rescaling was carried out, considering the somewhat reduced Tl bond strength [32] and the site occupation. The site occupation was found little too high, which may be to the higher scattering power of $Tl^{1+}$ compared to $Tl^{3+}$. The changes are recently summarized [33].

The complex chemical formula for the homologous series of $Tl^{3+}$-based superconductors may be written as

$$Tl^{3+}{}_{m-f-x}Tl^{1+}{}_{f}M^{s+}{}_{x}(Ba,Sr)_2(Ca_{1-y}Tl^{3+}{}_{y})_{n-1}Cu_nO_{m+2n+2-\delta}, n = 1,2,3,...$$

$m = 1$: TlO mono-layers (space group P4/mmm, $m = 2$: TlO bi-layers (space group I4/mmm). The space group notation is that of the 'averaged' structures.

**Table 1**. Calculation of the hole concentration $h^+$, the mean cationic charge $<q_c>$, and the critical length $\zeta$, using own crystal-chemical data of Tl-based cuprates as examples.

| Phase symbol | Tl-2201 | Tl-2212 | Tl-2223 | Tl-1223 |
|---|---|---|---|---|
| $a$ (Å) | 3.8656(3) | 3.8565(4) | 3.8498(4) | 3.848(4) |
| $c$ (Å) | 23.2247(18) | 29.326(3) | 35.638(4) | 15.890(10) |
| n | 1 | 2 | 3 | 3 |
| $f$ | 0 | 0 | 0 | 0.092 |
| $x$ | 0.11 | 0.31 | 0.42 | 0 |
| $y$ | - | 0.10 | 0.07 | 0.069 |
| $\delta$ | 0 | 0 | 0 | 0.18 |
| $h^+$ | 0.220 | 0.52 | 0.700 | 0.686 |
| $h^+/n$ | 0.220 | 0.26 | 0.233 | 0.229 |
| $<q_c>$ | 2.353 | 2.211 | 2.144 | 2.12 |
| $\zeta$ (Å) Ba-O $\parallel c$ | 1.945 | 2.014 | 2.010 | 2.073 |
| $T_{co}= 2740/<q_c>^4$ | 88.9 | 114.6 | 130 | 136 |
| $T_{co}= 1247.5 \cdot \sqrt{h^+}/(a \cdot \zeta)$ | 77.8 | 115.8 | 135 | 130 |
| $T_c$ (K) measured | 80 | 110 | 130 | 133 |
| Refs. | [29] | [29] | [28] | [30] [33] |

The hole concentration $h^+$ then yields

$$h^+ = 2-m+2f+\{x(3-s)-(n-1)y-2\delta\}.$$

Subtracting $h^+$ from the total charge of oxygen, taken as $O^{2-}$, the total cationic charge results. The mean cationic charge is calculated by division of the number of cations as

$$<q_c> = 2(m+2n+2-\delta)-h^+/(m+2n+1).$$

The amount of lethally toxic $Tl^{1+}$ besides $Tl^{3+}$ may easily be replaced by the environmentally benign $Rb^{1+}$ or $Cs^{1+}$ cations of comparable size (Figure 2), if you do not even use the $^{137}Cs$ isotope, without alter the charge balance. However, further replacement of $Tl^{3+}$ and $Ba^{2+}$, respectively, by these alkali earth cations would reduce the mean cationic charge $<q>_c$ and, according to the Eq. (1), should enhance $T_c$. Such replacement may be further supported by substitution of some oxygen by group VII anions like $F^{1-}$ or more polarizable ones such as $Br^{1-}$ or even $I^{1-}$. Fluorine substitution was already successfully applied in Bi-based superconductors [34], and in Hg-based compounds under high pressure with the highest $T_{co}$ known [35] [36]. Fluorine substitution causes shrinkage of the lattice parameters and enhances $T_c$ to its optimum due to internal 'chemical' pressure. In Table 3 some $Cs^{1+}$ and $F^{1-}$ substitutions are proposed assuming the possible adaption of a stable perovskite-related crystal structure.

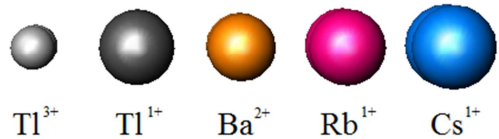

**Figure 2**. Relative size of cations, if coordinated by six anions.
Cation radii in Å: $Tl^{3+}$ 0.89, $Tl^{1+}$ 1.50, $Ba^{2+}$ 1.35, $Rb^{1+}$ 1.52, $Cs^{1+}$ 1.67.

**Table 2.** Calculation exercise of Cs$^{1+}$ substitution in the Tl-1223 model compound. The first elected substitution corresponds to the actual Tl-1223 composition, merely with a small amount of Cs$^{1+}$ instead of Tl$^{1+}$.

| Notation | Cations | | | | | | | Anions | |
|---|---|---|---|---|---|---|---|---|---|
| Formula subscripts | 1 | | 2 | | 2 | | 3 | 9 | |
| Elements | Tl$^{3+}$ | Cs$^{1+}$ | Ba$^{2+}$ | Cs$^{1+}$ | Ca$^{2+}$ | Tl$^{3+}$ | Cu$^{2+}$ | O$^{2-}$ | F$^{1-}$ |
| Substitution | 0.908 | 0.092 | 2 | - | 1.862 | 0.138 | 3 | 8.82 | - |
| Charge | +2.724 | 0.092 | +4 | - | +3.724 | +0.414 | +6 | -17.64 | - |
| $h^+ = 0.686$ | | | $<q>_c = 2.12$ | | | | $T_{co} = 136$ K | | |
| Substitution | 0.75 | 0.25 | 2 | - | 1.862 | 0.138 | 3 | 8.33 | 0.67 |
| Charge | +2.25 | +0.25 | +4 | - | +3.724 | +0.414 | +6 | -16.6 | -0.7 |
| $h^+ = 0.692$ | | | $<q>_c = 2.080$ | | | | $T_{co} = 146$ K | | |
| Substitution | 0.5 | 0.5 | 2 | - | 1.862 | 0.138 | 3 | 7.82 | 1.18 |
| Charge | +1.5 | +0.5 | +4 | - | +3.724 | +0.414 | +6 | -15.64 | -1.18 |
| $h^+ = 0.682$ | | | $<q>_c = 2.017$ | | | | $T_{co} = 166$ K | | |
| Substitution | - | 1 | 0.9 | 1.1 | 2 | - | 3 | 5.59 | 3.41 |
| Charge | - | +1 | +1.8 | +1.1 | +4 | - | +6 | -11.18 | -3.41 |
| $h^+ = 0.690$ | | | $<q>_c = 1.738$ | | | | $T_{co} = 300$ K | | |

A Cs-1223 compound of the formula Cs(Cs$_{1.1}$Ba$_{0.9}$)Ca$_2$Cu$_3$O$_{5.6}$F$_{3.4}$ with an optimum hole concentration of $h^+ \approx$ 0.7 = 3·0.233 and a mean cationic charge of $<q>_c = 1.738$ could yield a transition temperature near $T_{co} = 300$ K. The effect of crystal lattice swelling due to large cation substitution may be compensated by a larger permittivity of the compound (see Eq. 3).

Recently I proposed an Cs-elpasolite as candidate for solar cell application, namely Cs$_2$(Na,Cu,Ag)$_1$Bi$_1$(I,Br)$_6$ [14]. What would happen by hole doping of such compound, for instance by substitution of I$^{1-}$ or Br$^{1-}$ by some O$^{2-}$? A chemical content near Cs$_2$(Na,Cu,Ag)$_1$Bi$_1$(I,Br)$_{5.4}$(O,S)$_{0.6}$ looks very interesting, though it shows an optimum $h^+$ concentration and would reach a $T_c \approx 370$ K, counting the formally enhanced oxidation state of Cu or Ag, respectively. Alternative sulfide replacement is considered owing to the well adapted ionic radius near that of I$^{1-}$ or Br$^{1-}$. A vision is that both properties, high solar efficiency and superconductivity, occur together in slightly altered parts of a single compound and can be shared in future.

## 2.2.2 Nano-Scale Sandwich Structures Based on Tenorite (CuO) Layers

The discovery of multiferroic properties of slightly acentric CuO (tenorite) [37] as well as the formation of elongated rocksalt-type T-CuO thin films onto a (100)-SrTiO$_3$ substrate [38] has increased immensely the interest in these compounds. At low temperature tenorite undergoes two successive magnetic transitions at $T_{N1} = $ 213 K and $T_{N2} = 230$ K [37] [39]. Ferroelectric response is indicated at $T_c = 230$ K together with the development of a spiral-magnetic order and spontaneous electric polarization $P_s$ along the *b* axis. Removing of the glide mirror plane would lead to the acentric point group 2. It is natural to identify the temperature of $T_{N2} = $ 230 K with the 220 K temperature found for filamentary superconductivity on tenorite samples. However, the intrinsically physical properties and the high permittivity near the phase transition may also support a real superconducting transition, if applied strain would enhance the small concentration of holes that already exists. Notably, the smallest observed Cu-Cu distance of 2.900 Å (Table A1) indicates metallic bonding behavior with the ability to create holes.

In tenorite and T-CuO, respectively, are twice as many copper atoms in the atomic layers compared to the CuO$_2$ nets, and the copper to copper distances are similar to the oxygen-oxygen distances of about 2.73 Å. Both nets are compared in Figure 3. T-CuO nets should be investigated in detail to unravel their possible suitability as building units for superconductors. Especially (100) oriented composite structures composed of tenorite (CuO), cuprite (Cu$_2$O) and metallic AuCu alloy supports will be investigated. In addition, also CuBr or CuI are considered as material using [001] as stack direction.

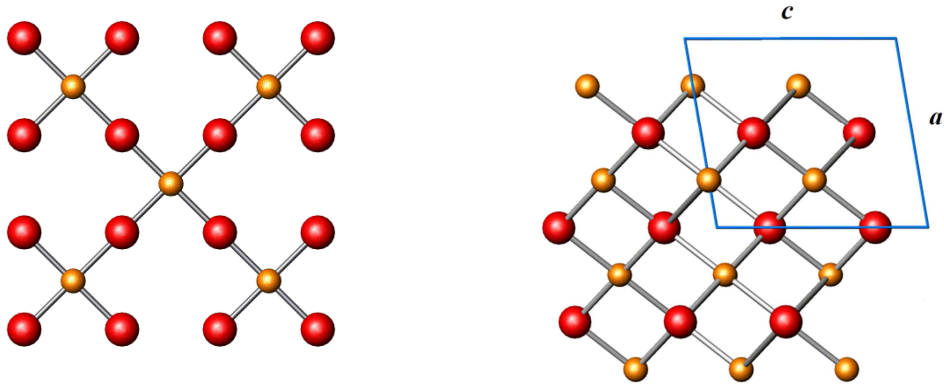

**Figure 3**. Left: $CuO_2$ net of high-$T_c$ superconductors. Right: Monoclinic crystal structure of tenorite (CuO), projected down [010] with blue outlined unit cell, copper atoms brownish, oxygen atoms red. The [101] and [10$\bar{1}$] directions, respectively, corresponding to 45° rotation, would yield the rocksalt lattice orientation.

The monoclinic crystal structure of tenorite is indeed a strongly folded and distorted rocksalt structure formed by unit cell twinning along (010) planes (Figure 4).

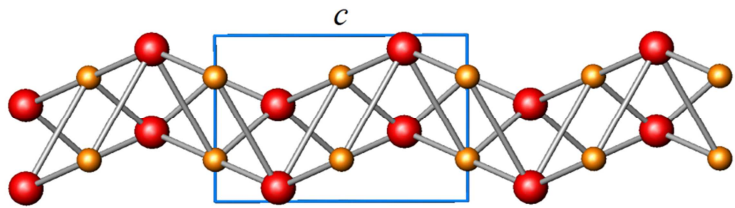

**Figure 4**. Crystal structure of tenorite, projected onto (100). The drawing clearly indicated that the monoclinic structure is actually a twin-folded and distorted rocksalt structure as already proposed in [10].

Twinning of the rocksalt structure type was first described in the system $PbS-Bi_2S_3$ by the present author and may serve as an example for CuO twinning too, although it does not be 'chemical' twinning, because contrary to the $PbS-Bi_2S_3$ system, the chemical content remains unchanged [40] [41].
Starting with the T-CuO structure, twin-folding to tenorite may be initiated by local $Cu^{1+}$–$Cu^{3+}$ charge disproportionation, which would remove the *Jahn-Teller* distortion of $Cu^{2+}$ caused by its $d^9$ electron configuration [42]. Interestingly on this aspect is that the *dd*-excitations of tenorite, characterized by its *UV-VIS* spectrum as well as *EPR* studies, have been interpreted on the basic of a $d^2$ configuration rather than a $d^9$ configuration [43]. This suggests dimerization of formally 2+ charged Cu pairs, which may actually considered as $Cu^{1+}$-$Cu^{3+}$ pairs. Not surprisingly, a very small quantity of $Cu^{3+}$ was found as relict in tenorite [44]. The few attributed holes order themselves in charge stripes along [100] [45]. The hole carriers are obviously sourced out into the (010) twin plane sinks, extending down [100] (see Figure 4).
Turning now to copper oxide composites, room temperature superconductivity is proposed for a composition near $6CuO·Cu_2O$ with $<q>_c \approx 1.75$. A (100)-$SrTiO_3$ substrate may be first coated with an AuCu I thin film, onto which a tenorite layer is deposited along [010], succeeded by a (100)-cuprite layer, and finished by a further tenorite layer in compliance with the stoichiometric specification. The layer sequence may be changed to deposit tenorite between cuprite layers, which can give a distinct result. The [101] direction of tenorite is sought to be oriented parallel to [100] of cuprite. The oxygen atoms on the interface of the tenorite layer will coordinate with copper atoms of the cuprite layer and vice versa (see Figure 5).
An open question is, whether self-doping is possible. If cuprite is considered as reservoir layer, one find short layer distances between copper layers and oxygen layers of $c/4 = 4.2698/4$ Å = 1.0674 Å, whereas the oxygen distances in cuprite are extreme large (Table 2). Cu-Cu bonding and d-orbital holes were observed in cuprite, and the four-coordinated oxygen atom is well 2- charged [46]. The copper atoms on the interface of both structures, which are initially 2+ and 1+ charged, respectively, can suffer a formal charge alteration, leading to doping with a sufficient amount of holes. Two processes can deliver holes:

1. Cu-Cu metallic bonding    $Cu^{2+} + Cu^{2+} \rightarrow (Cu\text{-}Cu)^{2+} + 2h^+$ (formally)
2. Charge disproportionation  $6CuO \cdot Cu_2O \rightarrow Cu_2O_3 \cdot 2CuO \cdot 2Cu_2O \rightarrow 4CuO \cdot 2Cu_2O + 2h^+$

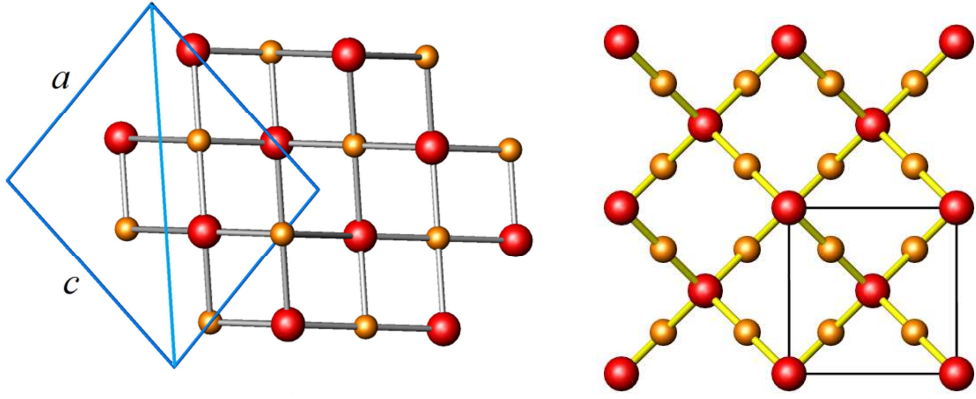

**Figure 5.** Illustration of the oxygen lattice match between tenorite (left) and cuprite (right), both depicted as projections onto (010). The light blue line in the tenorite picture indicates the [101] direction.

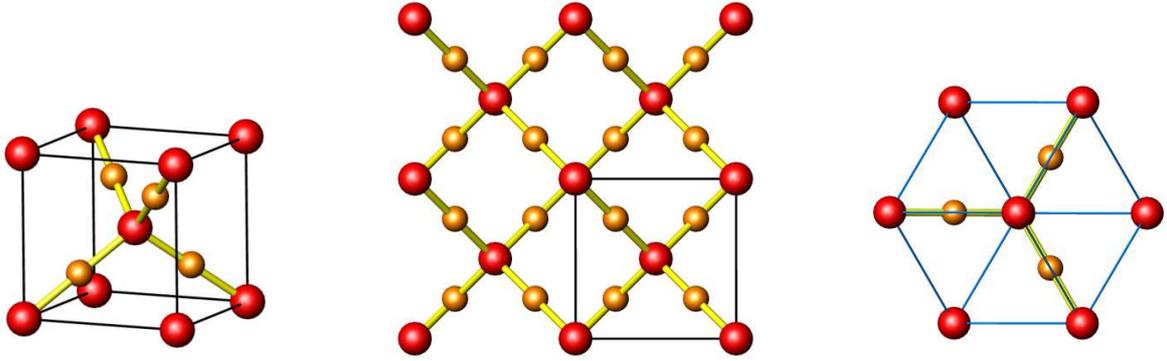

**Figure 6**. Left: Cubic crystal structure of cuprite ($Cu_2O$), depicting the characteristic twofold (dumbbell-like) oxygen coordination of $Cu^{1+}$. Middle: projection down [100]. Right: projection down [111].

The composition is chosen to satisfy the $<q>_c = 1.75$ condition for room temperature critical temperature. Strain may well favor charge disproportionation. One can consider the twin planes, depicted in Figure 4, as domain walls with soaked holes, showing a wall distance of $d_{(002)} = 2.5289$ Å. Using these geometrical specifications, hole stripes along [100] and domain extension along [001], respectively, then the [010] direction remains as direction for layer by layer deposition of CuO. If tenorite with its monoclinic angle of $β = 99.54°$ is deposited with the *ac* plane onto a tetragonal substrate such as an AuCu alloy, strain can cause multiple twinning along two planes, namely (100) and (001). In this way an incomplete and distorted *Penrose*-like parquet segment can be formed. In the empty space, corresponding to the acute Penrose tiles, dangling bonds and charge alteration of oxygen, respectively, can easily deliver holes.

*Harshman, Fiory and Dow* [15] derived another scaling relation to determine the critical temperature of optimum doped superconductors, which exemplarily will be applied to cross check the $T_{co}$ results (Table 4). The *HFD* rule gives

$$T_{co} = 1247.5 \cdot \sqrt{h^+/(d \cdot \zeta)} \text{ (K)}, \qquad (4)$$

where $h^+$ is the hole carrier concentration, $d$ (Å) the lattice parameter (Cu-Cu distance in the *ab* plane) and $\zeta$ (Å) represents a critical length down the charge reservoir layer, typically the Ba-O distance projected down [001]. This distance roughly corresponds to the *c*-axis coherence length $\xi_c$. Interestingly, with the change from $CuO_2$ nets to CuO ones the value of $d$ is reduced by a factor of √2, and correspondingly $T_{co}$ would be enhanced by this factor, if the other parameters remain unchanged.

Identifying the lattice plane spacing $d_{(004)} = 1.0674$ Å of cuprite as the critical length $\xi$ of a 'charge reservoir layer', then the *HFD* rule can be applied. Table 4 summarizes results for specific values chosen, varying the hole concentration and the critical length $\xi$.

**Table** 4. Rough calculation of $T_{co} = 1247.5 \sqrt{(nh^+)/(d \cdot \xi)}$ according to the *HFD* rule.

| Domain extension $d$ (Å) Tenorite | Critical length $\xi$ (Å) Cuprite | Hole concentration $h^+$ | $T_{co}$ (K) |
|---|---|---|---|
| $d_{(002)} = 2.5289$ | $d_{(004)} = 1.0674$ | 0.228 | 221 |
| | | 2 · 0.228 | 312 |
| $d_{(202)} = 1.5803$ | $d_{(222)} = 1.2325$ | 0.228 | 306 |
| | | 2 · 0.228 | 433 |

Surprisingly, the mysterious critical temperatures of previous experiments, namely 220 K [3] [4] and about 300 K [5], could be reproduced, thereby explaining their $T_{co}$ ratio of about √2. Numerically, room temperature superconductivity seems possible, if these specifications can be realized in practice. Notably, only two copper-oxygen layers are considered to spend optimum holes.

Alternatively, other orientations of thin film deposition may be tested. For instance, (001)-layers of Au or AuCu can be deposited on a cleaved mica sheet, succeeded by a (111)-oriented cuprite layer (see Figure 4, right). Then tenorite will be deposited with pseudo-hexagonal layers oriented as in Figure 7, followed by another cuprite film. In this way one avoids that not wanted T-CuO forms at all.

Thus, two possible orientations for epitaxial layer growth of copper oxides are indicated as consequence of this modeling: along (010) of tenorite and along (111) of cuprite, respectively. For an elaborated experiment, also tenorite single crystals could be cut in the wanted orientation. Under less controlled conditions, the deposition of (010)-oriented layers likely happens with unfortunate domain extension normal to the film surface. The extremely short filaments would lead to unstable current transport.

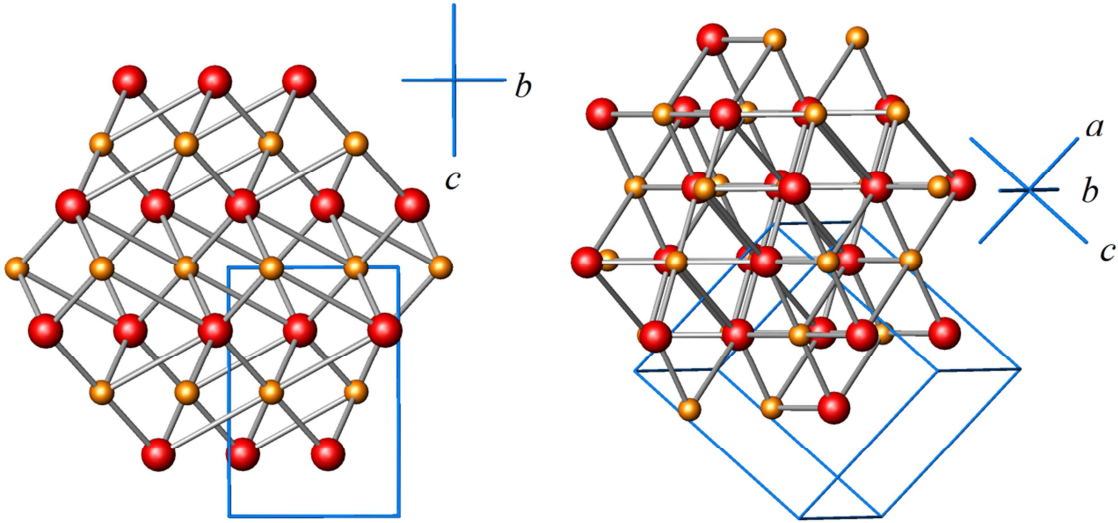

**Figure 7**. Two selected orientations of tenorite. Left: Atomic layers perpendicular to [001] with layer spacings of $d_{(004)} = 1.2645$ Å. Right: A tenorite orientation showing pseudo-hexagonal nets as well as atom layers with the lattice plane spacing of $d_{(202)} = 1.5803$ Å perpendicular to the [101] direction (along the paper longside).

Tetragonal CuO (siemonsite), which may initially be formed, is not wanted because of its large layer separation of $c$ = 5.30 Å. Tetragonal CuO should be given the opportunity to collapse into the 'dying swan' structure of tenorite by varying the coating thickness. Also paramelaconite, $Cu_4O_{3-x}$, is not considered as superconductor material [47]. The data in Table A1 were reported for purpose of completeness.

Finally, attention needs to be given to nano-structures that may be built up from tenorite and CuI (or CuBr, CuCl), respectively. Adequate doping implied, a composition of 2CuO·CuI with a *Fibonnaci* mean cationic quotient of $<q>_c$ = 5/3 would be able to reach a transition temperature of $T_c \approx$ 355 K. As pursued recently for solar cell application [14], a (111) oriented AuCu alloy support is recommended to deposit (111)-CuI first and then tenorite.

The hypothetical compound $BaCuO_2$ with puckered T-CuO nets [10] may be deposited onto a (100)-AuCu substrate, itself deposited onto (100)-$SrTiO_3$. $Cs^{1+}$ substitution for $Ba^{2+}$ could deliver holes. The mean cationic charge for a composition of $Ba_{0.772}Cs_{0.228}CuO_2$ would yield $<q>_c$ = 1.772, leading to $T_{co} \approx$ 278 K for the assumed critical temperature. However, the critical length $\xi$ down the 'charge reservoir' would be too large to uphold this result by an alternative calculation, using the *HFD* rule [15].

In conclusion, there are strong arguments for a reappraisal of tenorite-cuprite as well as tenorite-CuI (CuBr, CuCl) sandwich structures as candidates for room temperature superconductivity.

### 2.2.3 An Anti-Perovskite Option

Based on the recent investigation of the crystal structure of the $H_2S$ based high-pressure superconductor by *Gordon et al.* [2] the idea ripened to construct a hypothetical anti-perovskite structure, which would fulfill the conditions of low cationic charge. Many years ago, the present author investigated the monoclinic crystal structure of the ferroic compound $Pb_3GeO_5$ and presented it as anti-perovskite structure type, where the large $(GeO_4)^{4-}$ units occupy the A-sites and $(OPb_3)^{4+}$ the B-site of a perovskite type [48]. According to the investigation of *Gordon et al.* [2] the superconducting $H_2S$ modification may as well be understood as anti-perovskite with $(SH)^-$ occupying the A-site and $(SH_3)^+$ the B-site, respectively.

The $Pb_3GeO_5$ lead germanate is ferroelastic with a pronounced domain structure. Domains can easily be restored in the other orientation states. When hole doping would be possible, nesting of the charge carriers in the ferroelasic domain wall sinks could be proposed. Crystal-chemically interesting would be a composition of $(SiO_2F_2)^{2-}(BrRb_3)^{2+}$ with a mean cationic charge of $<q>_c$ = 1.75, leading to a possible transition temperature of $T_{co} \approx$ 292 K, if optimal doped. The realization of such compound is doubtful. However, high pressure may be a mean to outsmart limiting factors.

### 3. Conclusions

From an empirical relation between the critical temperature $T_{co}$ of optimum doped superconductors and the mean cationic charge $<q>_c$ new insights were obtained, strongly indicating the fractal character of high-$T_c$ superconductivity. The optimum hole concentration of $h^+$ = 0.229 can be linked with the universal fractal constant of $\delta_1$ = 8.72109 characteristic for the renormalized *Hénon* map, and the width of superconducting domains is governed by *Fibonacci* numbers, including mixed domain states (applying the mean of consecutive numbers). Experiment and theory in this field of science may be steered into new paths. There is evidence to favor low $<q>_c$ compounds with high permittivity to reach superconductivity above room temperature. The simple relation $T_{co}(K) = 2740/<q_c>^4$ points to compounds around $<q>_c \approx$ 1.74 to reach this dream. Some compounds are proposed for experiments. Besides $Cs^{1+}$ substitution in the known Tl-1223 cuprate, especially composites consisting of multiferroic tenorite and cuprite layers, respectively, were recommended because of previously reported filamentary superconductivity of such composites. Tenorite with twice as many copper atoms in the structural layers compared to $CuO_2$-based nets is highly interesting. Cesium is an element of the green future and has already found its way into smart perovskitic solar cells. Modeled on the anti-perovskitic high-pressure modification of $H_2S$, a $(SiO_2F_2)^{2-}(BrRb_3)^{2+}$ anti-perovskite may be a further option for high-$T_c$. Once again it should be stressed that a low cationic charge is optimal for both solar cells and superconductors.

# Appendix

## A1. Proper Substrates for Epitaxial Thin Film Growth

A frequently used substrate for epitaxial growth of thin films of superconductors is (100)-SrTiO$_3$. It can be used for the deposition of both types of copper oxide based nets. If one thinks about a metallic substrate, AuCu alloys would be best adapted to an epitaxial growth of any copper oxide composite. Au has the tendency to favor (111) layers, if deposited on smooth surfaces. However, deposition onto a (100)-SrTiO$_3$ would favor (100) growth. The cubic AuCu L1$_0$ phase with an atomic ratio of 1:1 has a lattice parameter of $a = 3.859$ Å. The atomic distance yields $a/\sqrt{2} = 2.729$ Å. In contrast, the tetragonal AuCu I phase has lattice parameters of $a = 3.942$ Å and $c = 3.670$ Å, whereas the AuCu II phase is orthorhombic with lattice parameters of $a = 3.946$ Å, $b = 3.957$ Å, and $c = 3.646$ Å, respectively [49]. The $c$ parameters of the non-cubic phases are similar to the cubic lattice parameter of pure Cu. As a result of these considerations one should choose the AuCu I alloy with its smaller layer separation down [001] of 3.67 Å, because a small layer distance may be important in case such metallic buffer would contribute to superconducting properties. A proper substrate for the deposition of thin films of Cu$_2$O (cuprite) is (100)-MgO due to its cubic lattice parameter of $a = 4.212$ Å compared to $a = 4.2696$ Å for cuprite.

If simply extruded foils of pure copper are used as a substrate for preliminary experiments, their annealing under protective hydrogen atmosphere at 600 °C provides a highly ductile product that does not break, even if most frequently bent.

## A2. Crystallographic data and physical properties for copper and copper oxides.

$\Theta_D$ *Debye* temperature, $v_s$ sound velocity, $\varepsilon$ permittivity, $\rho$ density, $T_N$ *Neel* temperature, $E_g$ energy gap.

| Phase | copper | tenorite | 'siemonsite' | cuprite | paramelaconite |
|---|---|---|---|---|---|
| Formula | Cu | CuO | T-CuO | Cu$_2$O | Cu$_4$O$_{3-x}$ |
| Space group | Fm3m | C2/c (293 K) |  | Pn3m | I4$_1$/amd |
| $a$ (Å) | 3.6147 | 4.6837 | 3.905 | 4.2696 | 5.837 |
| $b$ (Å) |  | 3.4226 |  |  |  |
| $c$ (Å) |  | 5.1288 | 5.30 |  | 9.932 |
| $\beta$ (°) |  | 99.54 |  |  |  |
| $V$ (Å$^3$) |  | 81.080 | 80.82 | 77.832 | 4 · 84.598 |
| $\rho$ (kg/m$^3$) | 8.937 | 6.516 | 6.54 | 6.106 | 5.93 |
| Cu–Cu (Å) | 2.5560 | 2.9005  3.0830  3.1733 | 2.761 | 3.0191 | 2.9185 |
| $\varepsilon$ |  | 5.9 ∥ [010]<br>6.2 ∥ [101]<br>7.8 ∥ [10$\bar{1}$] |  | 7.11 $\varepsilon(0)$<br>6.46 $\varepsilon(\infty)$ | 4.24 $\varepsilon(\infty)$ *) |
| $\Theta_D$ (K) | 347 (0 K)<br>310 (298 K) | 391 |  | 188 |  |
| $v_s$ (m/s) | 4.76·10$^3$ (long)<br>2.33·10$^3$ (trans) | 6.4·10$^3$ [100]<br>4.1·10$^3$ [010]<br>7.8·10$^3$ [001]<br>5.4·10$^3$ [101]<br>9.1·10$^3$ [10$\bar{1}$]<br>6.8·10$^3$ [111] |  |  |  |
| $T_N$ (K) |  | 213<br>230 |  |  | 45-55<br>120 |
| $E_g$ (eV) |  | 1.2 |  | 2.137 | 1.34 (indirect)<br>2.47 (direct) |

*) calculated from the refractivity index